\documentclass[aps,nofootinbib]{revtex4-1}
\pdfoutput=1
\usepackage{graphicx,subfigure}
\usepackage{amsmath,amssymb}
\usepackage{xcolor}
\usepackage{mathrsfs}  
\usepackage{ulem}

\def\beq{\begin{equation}}
\def\eeq{\end{equation}}
\def\beqr{\begin{eqnarray}}
\def\eeqr{\end{eqnarray}}
\def\bdpm{\begin{displaymath}}
\def\edpm{\end{displaymath}}
\def\half{\frac{1}{2}}
\definecolor{lgray}{gray}{0.6}

\newcommand{\nnb}{\nonumber}

\newcommand{\tbt}{t_\beta}
\newcommand{\tth}{t_\theta}
\newcommand{\cth}{c_\theta}
\newcommand{\sth}{s_\theta}

\newcommand{\msbar}{\overline{\textrm{MS}}}
\newcommand{\Veff}{V_\textrm{eff}}

\begin{document}

\title{
Vacuum stability of conformally invariant scalar dark matter models
}

\author{Yeong Gyun Kim}
\email{ygkim@gnue.ac.kr}

\affiliation{ 
	Department of Science Education, 
	Gwangju National University of Education, Gwangju 61204, Korea
}

\author{Kang Young Lee}
\email{kylee.phys@gnu.ac.kr}

\affiliation{ 
	Department of Physics Education \& RINS,
	Gyeongsang National University, Jinju 52828, Korea
}

\author{Jungil Lee}
\email{jungil@korea.ac.kr}

\author{Soo-hyeon Nam}
\email{glvnsh@gmail.com}

\affiliation{ 
Department of Physics, 
Korea University, Seoul 02841, Korea
}

\date{\today}

\begin{abstract}

We discuss vacuum structure and vacuum stability in classically scale-invariant renormalizable models 
with a scalar dark matter multiplet of global ${\cal O}(N)$ symmetry together with an electroweak singlet scalar mediator. 
Our conformally invariant scalar potential generates the electroweak symmetry breaking via the Coleman-Weinberg mechanism, 
and the new scalar singlet mediator acquires its mass through radiative corrections of the scalar dark matters 
as well as of the standard model particles. 
Taking into account the present collider bounds, we find the region of parameter space 
where the scalar potential is stable and all the massless couplings are perturbative up to the Planck scale.
With the obtained parameter sets satisfying the vacuum stability condition,
we present the allowed region of new physics parameters satisfying the recent measurement of relic abundance,
and predict the elastic scattering cross section of the new scalar multiplet 
into target nuclei for a direct detection of the dark matter.
We also discuss the collider signatures and future discovery potentials of the new scalars.

\end{abstract}

\maketitle

\section{Introduction}

Discovery of the Higgs boson at the CERN Large Hadron Collider (LHC)
has closed the particle contents of the standard model (SM).
However, there is no room for the nonbaryonic cold dark matter (DM) 
of the Universe in the SM Lagrangian. 
The existence of the DM is strongly supported 
by a variety of astrophysical observations 
and cosmological implications in the early Universe.
Stil, no evidence of the DM was obtained in the direct detection 
and high-energy collider experiments.
Thus, it is well motivated to consider theoretical models 
in which the DM candidates are included in the separate hidden sector 
connected to the SM with small couplings through some mediators.

Recently, the precise measurement of the Higgs boson mass, 
$m_H = 125.25 \pm 0.17$ \cite{ParticleDataGroup:2020ssz},
allows us to study the SM vacuum structure in more detail.
The present values of $m_H$, $m_t$ and $\alpha_s(M_Z)$ 
imply the metastability of the electroweak vacuum in the SM,
which is brought by the Higgs quartic coupling 
turning negative at some high-energy scale.
The vacuum stability issue has been investigated by many authors
\cite{Ghosh:2017fmr,Beniwal:2015sdl,Oda:2015gna,Branchina:2014rva,Khan:2014kba,Branchina:2014usa,Baek:2012uj,
	Masina:2012tz,Alekhin:2012py,Degrassi:2012ry,Bezrukov:2012sa,Lebedev:2012zw,Elias-Miro:2011sqh}.
Additional scalars are often introduced to rescue the vacuum from metastability,
and those can be mediator fields and/or the DM candidates as hidden sector particles.

In this work, we adopt the renormalizable DM model 
with the scale invariance at classical level proposed in our previous work \cite{Jung:2019dog}
in order to preserve the naturalness of the model up to the Planck scale $M_P$.
Softly broken scale invariance could be a solution to the hierarchy problem \cite{Bardeen,Shaposhnikov:2018nnm}.
Being assigned the scale invariance at classical level, 
dimensionful terms do not exist in the model Lagrangian 
but are generated by the quantum corrections 
to achieve the electroweak symmetry breaking (EWSB)
as done by Coleman and Weinberg \cite{Coleman:1973jx}.
In this model, the DM candidate is a scalar multiplet of global ${\cal O}(N)$ symmetry,
and the mediator field is a singlet scalar. 
For a systematic minimization of the effective potential, 
we find the flat direction of the scalar potential
and obtain the one-loop radiative corrections to lift up the flat direction potential 
following the Gildener and Weinberg (GW) formalism \cite{Gildener:1976ih}.
Then we get the local minimum of the potential 
to generate the scalar mass terms which give rise to the EWSB. 
As a result of conformal symmetry breaking, 
a light pseudo-Nambu-Goldstone boson appears 
and is mixed with the SM-like Higgs boson emerging 
from two light scalar particles in this model.
We assign ${\cal O}(N)$ global symmetry to the DM scalars 
for the stability of the hidden sector.
A similar model was studied for $N=2$ case 
in Refs. \cite{Ghorbani:2015xvz,Ghorbani:2017lyk},
but they simply assumed that the SM-DM coupling($\lambda_{h\phi}$) is zero
in order to decouple the DM sector from the SM Higgs.
In general, however, 
such an interaction term is not forbidden 
by a discrete symmetry such as $Z_2$ symmetry theoretically
and also is very important to explain 
the current astronomical observables phenomenologically, 
as discussed in our earlier work.

In addition to the phenomenological study,  
we investigate the vacuum stability in this model
and show that the nonzero $\lambda_{h\phi}$ 
plays an important role in stabilizing the scalar potential as well. 
The Higgs quartic coupling evolves 
with additional positive contributions from the new scalar quartic couplings in the dark sector 
and thus the metastability of the electroweak vacuum can be cured.
Besides $\lambda_{h\phi}$, 
the DM self-coupling $\lambda_{\phi}$ is also important in stabilizing the scalar potential. 
Although its direct contributions to the DM phenomenology were discarded in our previous study, 
its contribution can alter other parameters at one-loop level. 
Therefore, we newly consider the nonzero contributions of $\lambda_{\phi}$ in this work.
Taking into account all of the theoretical consideration and new parameters mentioned, 
we demand the vacuum stability as a new theoretical constraint to study the DM and collider phenomenology.
We find the parameter sets with the stable vacua 
which satisfy the relic abundance as well as the direct detection bounds.
Based on the obtained vacuum stability condition, 
we get the conservative bounds on the mixing angle between neutral scalar bosons
 together with the DM mass.
The future prospect of the allowed parameter set to test in the collider phenomenology is also discussed.

The outline of this paper is as follows.
In Sec.~II, we describe the model and discuss the physical degrees of freedom and model parameters
together with the one-loop effective potential.
We show the the beta functions of the couplings and discuss the vacuum stability conditions in Sec.~III.
In Sec.~IV, the relic density and the direct detection limits for the DM are shown. 
The implication of the collider phenomenology is also discussed.
Section V summaries the results and concludes.

\section{models}

Our model was discussed in the earlier study in detail \cite{Jung:2019dog},  so we can be brief.	
We consider a scalar multiplet $\phi = (\phi_1, \cdots, \phi_N)^T$, 
which is the fundamental representation of a global ${\cal O}(N)$ group as a DM candidate.
We also introduce a real scalar singlet $S$ as the mediator
which participates in the EWSB together with the SM Higgs doublet $H$. 
We demand the scale invariance at the classical level 
and the scalar potential consists of quartic interactions only as given by
\beq \label{eq:VS_potential}
 V(H, S, \phi) =  \lambda_h (H^{\dagger} H)^2 
             + \half\lambda_{hs} H^{\dagger} H S^2  
 	     + \half\lambda_{h\phi} H^{\dagger} H \phi^T\phi 
	     +\frac{1}{4}\lambda_{s\phi} S^2\phi^T\phi 
 	     + \frac{1}{4}\lambda_s S^4  
	     + \frac{1}{4}\lambda_\phi (\phi^T\phi)^2.
\eeq
Note that there exists an interaction term between the DM scalar $\phi$ and the SM Higgs $H$
with the nonzero coupling $\lambda_{h\phi}$,
while this coupling was discarded in similar models in Refs. \cite{Ghorbani:2015xvz, Ghorbani:2017lyk} as discussed earlier.
We should mention that there is no reason to forbid this term 
and have shown that $\lambda_{h\phi}$ plays an important role 
for the current astronomical phenomenology in our previous study.  
Moreover we will show that nonzero $\lambda_{h\phi}$ is essential for stabilizing the scalar potential in this paper. 
Also, the DM self-coupling $\lambda_{\phi}$ affects the beta functions of other couplings 
and does eventually alter the bounds of other parameters in one-loop level.
This was not discussed in the previous study, but will be shown in the next sections more clearly.
Alternatively to the above choice of the scalar structure,
one may combine $S$ with $\phi$ to obtain a fundamental of a global ${\cal O}(N+1)$
as studied in Refs.~\cite{Endo:2015ifa,Endo:2015nba}.
However, generating a proper Higgs mass requires a large enough mass of the scalar DM 
so that the Higgs-scalar couplings become too large. 
As a result, this kind of a simple setup makes the theory nonperturbative at a few TeV scale,
which undesirably ruins our original motivation to make our model valid up to the Planck scale.
That is why we resolved such an issue by separating the new scalar responsible for the EWSB from the DM sector.

To achieve the physical vacuum,
we will minimize the scalar potential of Eq.~(\ref{eq:VS_potential}) 
up to one-loop level. 
Following Gildener and Weinberg \cite{Gildener:1976ih},
first we minimize the tree level potential with the conditions
\beq
\frac{\partial V}{\partial H} |_{\langle H^0\rangle=v_h/\sqrt{2}} 
= \frac{\partial V}{\partial S} |_{\langle S\rangle=v_s}=0,
\eeq
which gives the following relations at some scale $\Lambda$,
\beq \label{eq:lambdamin}
\frac{\lambda_h(\Lambda)}{\lambda_s(\Lambda)} 
= \left( \frac{v_s}{v_h} \right)^4, ~~~
-\frac{2 \lambda_h(\Lambda)}{\lambda_{hs}(\Lambda)} 
= \left( \frac{v_s}{v_h} \right)^2,
\eeq
where the nonzero vacuum expectation values (VEVs) can be developed
as $\langle H^0 \rangle=v_h/\sqrt{2}$ and $\langle S \rangle = v_s$ after EWSB.
We let $\tan\beta (\equiv \tbt ) = v_s/v_h$ hereafter. 
The minimization of the tree level potential performed at a particular scale $\Lambda$ gives a flat direction among the scalar VEVs.
Because of the $H-S$ mixing term, the quadratic terms of neutral scalar degrees of freedom $h$ and $s$, 
defined by $H^0=(v_h+h)/\sqrt{2}$ and $S=v_s+s$, arise even at the tree level,
and the mass matrix is written as
\beq \label{eq:mass_matrix}
M^2 = 2 \lambda_h v_h^2 \left( \begin{array}{cc} 
        1 &\ - 1 /\tbt
	\\[1pt]
        - 1 /\tbt &\ 1 /\tbt^2
       \end{array} \right) .
\eeq
The corresponding scalar mass eigenstates $h_1$ and $h_2$ are obtained as admixtures of $h$ and $s$,
\beq \label{eq:scalar_mixing}
\left( \begin{array}{c} h_1 \\[1pt] h_2 \end{array} \right) =
\left( \begin{array}{cc} \cos \theta &\ -\sin \theta \\[1pt]
	\sin \theta &\ \cos \theta \end{array} \right)
\left( \begin{array}{c} h \\[1pt] s \end{array} \right) ,
\eeq
with the mixing angle $\theta$.
After diagonalizing the mass matrix, we obtain $\tan \theta = - \tbt$ or $1/\tbt$.
The mixing angle $\theta$ should be very small due to the LEP constraints \cite{Barate:2003sz}.
Then, experimental constraints disfavor the case of $\tan \theta = - \tbt$ 
as similarly discussed in Refs. \cite{Farzinnia:2013pga, Farzinnia:2014yqa}.
Thus we only choose $\tan \theta (\equiv \tth) = 1/\tbt$ in this work.
As a result, $\lambda_{hs}$ and $\lambda_{s}$ are suppressed by $\tth^2$ and $\tth^4$, respectively,

At tree level, the physical masses of the two scalars ($h_1$, $h_2$) and the DM scalar $\phi$
are obtained as
\beq \label{eq:scalar_mass_tree}
M^2_1 = 2 \lambda_h v^2 \tth^2, \quad M^2_2 = 0, \quad
M_{\phi}^2 = \frac{v^2}{2}\left(\lambda_{h\phi}\sth^2 + \lambda_{s\phi}\cth^2\right),
\eeq
where  $\sth \equiv \sin\theta$, $\cth \equiv \cos\theta$, and
$v^2 = v_h^2 + v_s^2$ is considered to be the VEV of the radial component of a scalar field composed of $h$ and $s$. 
The value of $v$ is determined from the radiative corrections and is set to be the scale about $\Lambda$ according to GW.
From the above equation, one can find the massless mode exists.
We take $h_1$ to be the SM-like Higgs boson which has the tree level masses and 
the other scalar $h_2$ to be the massless mode.

At one-loop level, the radiative corrections lift up the flat direction
and generate the mass of the massless mode.  
Including the radiative corrections, we write the scalar effective potential as
\beq
\Veff(h_{1c},h_{2c}) = V^{(0)}(h_{1c},h_{2c}) + V^{(1)}(h_{1c},h_{2c}),
\eeq
where
\beqr
V^{(0)}(h_{1c},h_{2c}) &=& 
 \frac{\lambda_{h}}{4}\left(\cth h_{1c} - \sth h_{2c}\right)^4
+ \frac{\lambda_{s}}{4}\left(\sth h_{1c} + \cth h_{2c}\right)^4 
+ \frac{\lambda_{hs}}{4}\left(\cth h_{1c} - \sth h_{2c}\right)^2
\left(\sth h_{1c} + \cth h_{2c}\right)^2 \nnb \\[1pt]	
V^{(1)}(h_{1c},h_{2c}) &=& 
\sum_P n_P \frac{\bar{m}_P^4(h_{1c},h_{2c})}{64\pi^2}
\left(\ln\frac{\bar{m}_P^2(h_{ic},h_{2c})}{\mu^2} - c_P\right),
\eeqr
$h_{ic}$ is the background value of the physical scalar $h_i$,
and $c_P = 3/2\ (5/6)$ for scalars and fermions (gauge bosons) 
in the $\msbar$ scheme.
The effective potential is computed at the renormalization scale $\mu$,
and $\bar{m}_P$ represent a field-dependent mass of fluctuating fields, $P = h_{1,2}, Z, W^\pm, t, \phi_i$.
Their degrees of freedom, $n_P$, are given by
\beq
n_{h_1}=n_{h_2}=n_{\phi_i}=1,\quad 
n_Z=3,\quad n_{W^\pm}=6, \quad n_t=-12.
\eeq
The field-dependent masses $\bar{m}_P$ are obtained 
in terms of $\tth \equiv \tan \theta$ as
\beqr \label{eq:eff_mass}
\bar{m}_{h_1}^2(h_{1c},h_{2c}) &=&
 \left(\lambda_h + \lambda_s\tth^4 + \lambda_{hs}\tth^2\right)\frac{3h_{1c}^2}{(1+\tth^2)^2}
+ \left[3\left(\lambda_h + \lambda_s\right)\tth^2 
+ \frac{\lambda_{hs}}{2}\left(1 - 4\tth^2 + \tth^4\right)\right]\frac{h_{2c}^2}{(1+\tth^2)^2}, 
\nonumber \\[1pt]
\bar{m}_{h_2}^2(h_{1c},h_{2c}) &=&
 \left[3\left(\lambda_h + \lambda_s\right)\tth^2 
+ \frac{\lambda_{hs}}{2}\left(1 - 4\tth^2 + \tth^4\right)\right]\frac{h_{1c}^2}{(1+\tth^2)^2}
+ \left(\lambda_h\tth^4 + \lambda_s + \lambda_{hs}\tth^2\right)\frac{3h_{2c}^2}{(1+\tth^2)^2}, 
\nonumber \\[1pt]
\bar{m}_{\phi_i}^2(h_{1c},h_{2c}) &=& 
 \left( \lambda_{h\phi} + \lambda_{s\phi}\tth^2 \right)\frac{h_{1c}^2}{2(1+\tth^2)}    	
+ \left( \lambda_{h\phi}\tth^2 + \lambda_{s\phi} \right)\frac{h_{2c}^2}{2(1+\tth^2)}, 
\nonumber \\[1pt]
\bar{m}_{Z}^2(h_{1c},h_{2c}) &=& 
\frac{M_Z^2}{v_h^2}\frac{\left(h_{1c}^2 + \tth^2h_{2c}^2\right)}{1 + \tth^2}, \qquad
\bar{m}_{W^\pm}^2(h_{1c},h_{2c}) \ =\ 
\frac{M_W^2}{v_h^2}\frac{\left(h_{1c}^2 + \tth^2h_{2c}^2\right)}{1 + \tth^2}, \nonumber \\[1pt]
\bar{m}_{t}^2(h_{1c},h_{2c}) &=& 
\frac{M_t^2}{v_h^2}\frac{\left(h_{1c}^2 + \tth^2h_{2c}^2\right)}{1 + \tth^2}.
\eeqr
Note that we omitted the Goldstone boson contributions here 
because those become zero eventually at the GW scale $\Lambda$ along the flat direction \cite{Kim:2019ogz}. 

\begin{figure}[!hbt]
	\centering%
	\includegraphics[width=8.3cm]{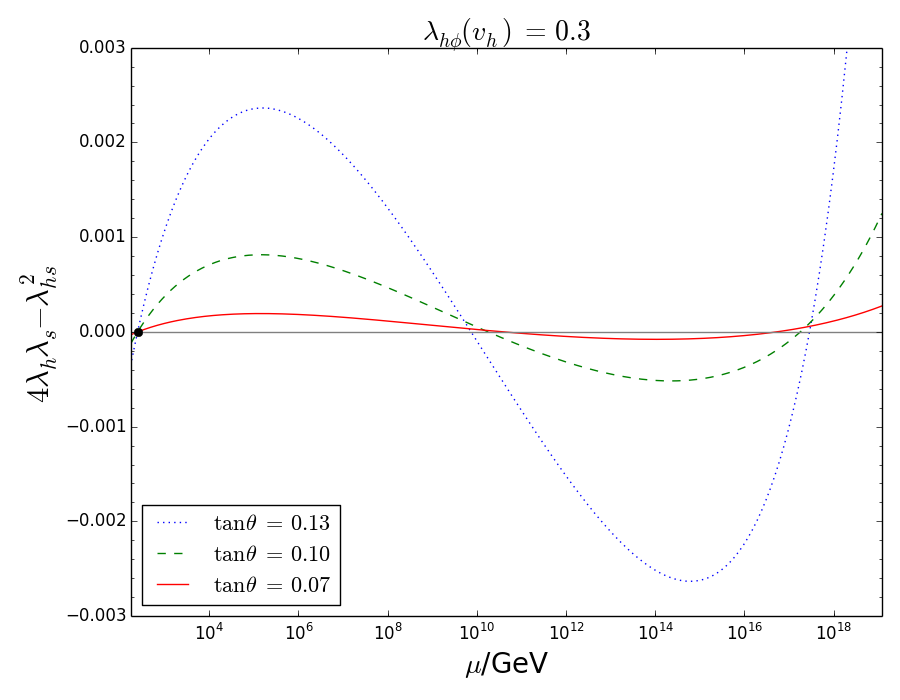} \quad  
	\includegraphics[width=8.3cm]{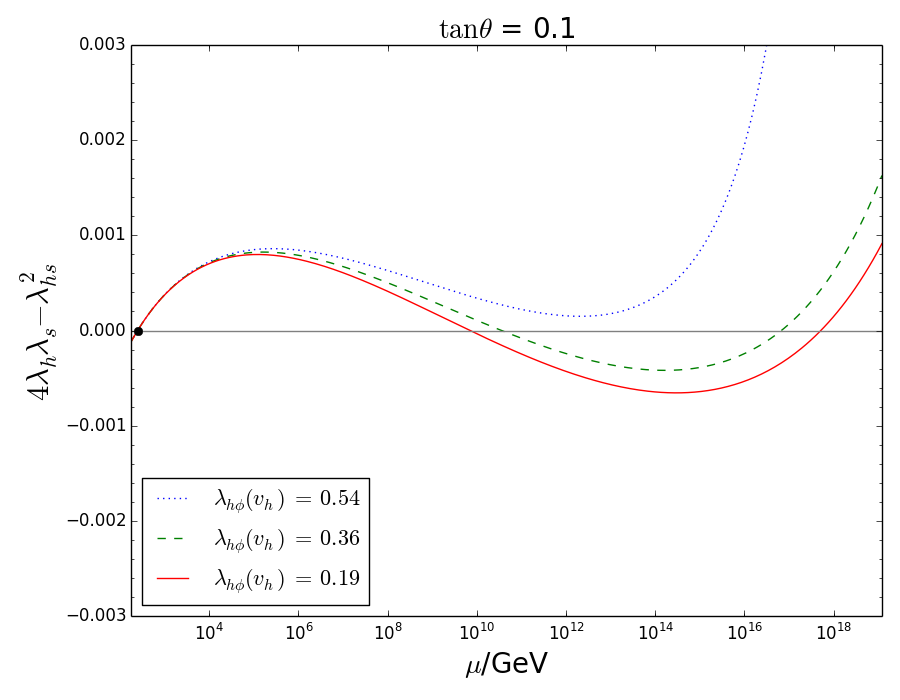}  	 
	\caption{ Behavior of the scalar coupling relation ($4\lambda_h\lambda_s -\lambda_{hs}^2$)
		for three different values of $\lambda_{h\phi}(v_h)$ and $\tan\theta$.
		The black dots represent the GW scale $\Lambda$
		at which the relation given in Eq.~(\ref{eq:lambdamin}) holds at one-loop level.} 
	\label{fig:gwcoupling}
\end{figure}

Because of the smallness of the mixing angle, 
running of the scalar couplings in Eq.~(\ref{eq:VS_potential}) is very slow with $\Lambda$,
and thus there exists some scale $\Lambda$ below the Plank scale $M_P \simeq 10^{19}$ GeV
at which Eq.~(\ref{eq:lambdamin}) is satisfied at one-loop level.
Similar discussions can be found also in Refs. \cite{Gildener:1976ih, Hempfling:1996ht}.  
Figure~\ref{fig:gwcoupling} depicts the scaling behavior of the coupling relation 
($4\lambda_h\lambda_s -\lambda_{hs}^2$)
by varying the Higgs-DM coupling $\lambda_{h\phi}(v_h)$ and the mixing angle $\tan\theta$.    
The running of the scalar couplings is obtained from the $\beta$ functions given in the next section.
One can see clearly from the figures that 
there are several scale points at which the following equation holds as in Eq.~(\ref{eq:lambdamin}): 
$4\lambda_h\lambda_s -\lambda_{hs}^2 = 0$. 
Among those scale points, 
we choose the lowest one (black dot) 
because it fits our purpose on the DM phenomenology.
The GW scale $\Lambda$ can be obtained 
by applying the minimization condition of the effective potential,
and we have $\Lambda \simeq 0.85 M_\phi $ for $N = 2$.
We will exploit the numerical values of the physical observables 
at this scale.	

Adopting the GW approach, we minimize the effective potential at $h_{1c} = 0$ and $h_{2c} = v$ 
keeping $h_1$ and $h_2$ to be the physical modes.
Then the field-dependent masses in Eq.~(\ref{eq:eff_mass}) 
can be further simplified at the GW scale by imposing the flat direction conditions 
in Eq.~(\ref{eq:lambdamin}) as 
\beqr \label{eq:eff_mass2}
\bar{m}_{h_1}^2(h_{2c}) &=& \frac{3}{4}\left[\lambda_h s_{2\theta}^2 + \lambda_s s_{2\theta}^2
+ \frac{\lambda_{hs}}{2}\left(\frac{1}{3} + c_{4\theta}\right)\right] h_{2c}^2
= 2\lambda_h\tth^2h_{2c}^2,  \quad 
\bar{m}_{h_2}^2(h_{2c}) = 3\left(\lambda_h\sth^4 + \lambda_s\cth^4 
+ \frac{1}{4}\lambda_{hs}s_{2\theta}^2\right)h_{2c}^2 = 0, \nonumber \\[1pt]
\bar{m}_{Z}^2(h_{2c}) &=& \frac{1}{4}\left(g_2^2+g_1^2\right)\sth^2h_{2c}^2
= M_Z^2\frac{h_{2c}^2}{v^2}, \quad
\bar{m}_{W^\pm}^2(h_{2c}) = \frac{1}{4}g_2^2\sth^2h_{2c}^2
= M_W^2\frac{h_{2c}^2}{v^2},	\nonumber \\[1pt]
\bar{m}_{t}^2(h_{2c}) &=& \frac{y_t^2}{2}\sth^2h_{2c}^2
= M_t^2\frac{h_{2c}^2}{v^2}, \quad
\bar{m}_{\phi_i}^2(h_{2c}) = \half\left( \lambda_{h\phi}\sth^2 + \lambda_{s\phi}\cth^2 \right)h_{2c}^2
= M_\phi^2\frac{h_{2c}^2}{v^2}. 
\eeqr
The one-loop masses of $h_1$ and $h_2$ are obtained from the effective potential 
\beqr \label{eq:radit_mass}
M_1^2 &=& \frac{\partial^2\Veff}{\partial h_{1c}^2}\Big{|}_{\substack{h_{1c}=0 \\ h_{2c}=v}} 
= 2 \lambda_h v^2 \tth^2, 	\nonumber \\[1pt]
M_2^2 &=& \frac{\partial^2\Veff}{\partial h_{2c}^2}\Big{|}_{\substack{h_{1c}=0 \\ h_{2c}=v}} 
= \frac{1}{8\pi^2v^2}\left(M_1^4 + 6M_W^4 + 3M_Z^4 -12M_t^4 + N M_{\phi}^4\right).
\eeqr 
We have to demand 
$N M_{\phi}^4 \ge 12M_t^4 - M_1^4 - 6M_W^4 - 3M_Z^4$ 
for $M_2^2 \ge 0$ and obtain that $M_{\phi} \gtrsim 265$ GeV for $N=2$.

In this study, we present the phenomenology of this model in terms of the following new physics (NP) parameters:
$M_{\phi}$, $\theta$, $\lambda_{h\phi}$, and $\lambda_{\phi}$.
The DM self-coupling $\lambda_\phi$ is irrelevant on the SM-DM interactions, so it was neglected 
in our previous study on the DM phenomenology \cite{Jung:2019dog}.
However it affects the RGE of the other couplings considerably as we will see in the next section.
Therefore, we probe the contribution of $\lambda_\phi$ to the vacuum stability 
and the corresponding DM phenomenology as well in this paper.  
The dependency of the model parameters at the GW scale $\Lambda$ are 
\beq \label{eq:parameters}
v = \frac{v_h}{\sth}, \quad v_s = \frac{v_h}{\tth}, \quad \lambda_h = \frac{M_1^2\cth^2}{2v_h^2}, 
\quad \lambda_{hs} = -\frac{M_1^2\sth^2}{v_h^2}, \quad \lambda_{s} = \frac{M_1^2\sth^2\tth^2}{2v_h^2},
\quad \lambda_{s\phi} = \left(\frac{2M_{\phi}^2}{v_h^2} - \lambda_{h\phi}\right)\tth^2.
\eeq
Given the fixed Higgs mass $M_1$ and $v_h \simeq 246$ GeV, 
we constrain the above four independent NP parameters 
by taking into account various theoretical considerations and experimental measurements in the following sections.

\section{Vacuum Stability }

\begin{figure}[!hb]
	\centering%
	\subfigure[$M_\phi = 1$ TeV, $\lambda_{h\phi}(\Lambda)$ = 0.3, $\lambda_\phi(\Lambda)$ = 0.01]
	{\label{fig:tan_theta} %
		\includegraphics[width=8.3cm]{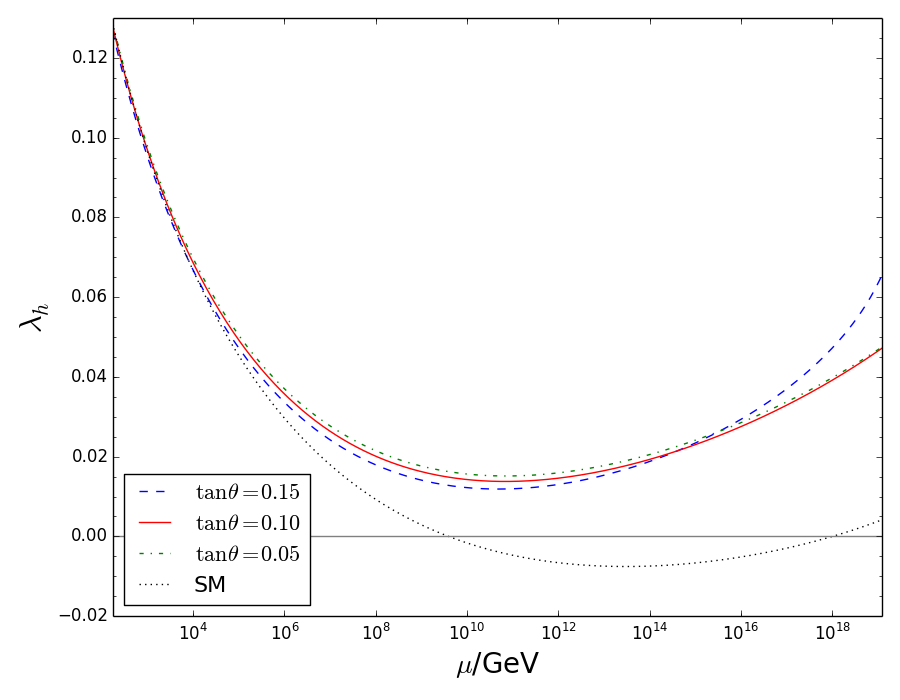}} \quad
	\subfigure[$\tan\theta = 0.1$, $\lambda_{h\phi}(\Lambda)$ = 0.3, $\lambda_\phi(\Lambda)$ = 0.01]
	{\label{fig:dm_mass} %
		\includegraphics[width=8.3cm]{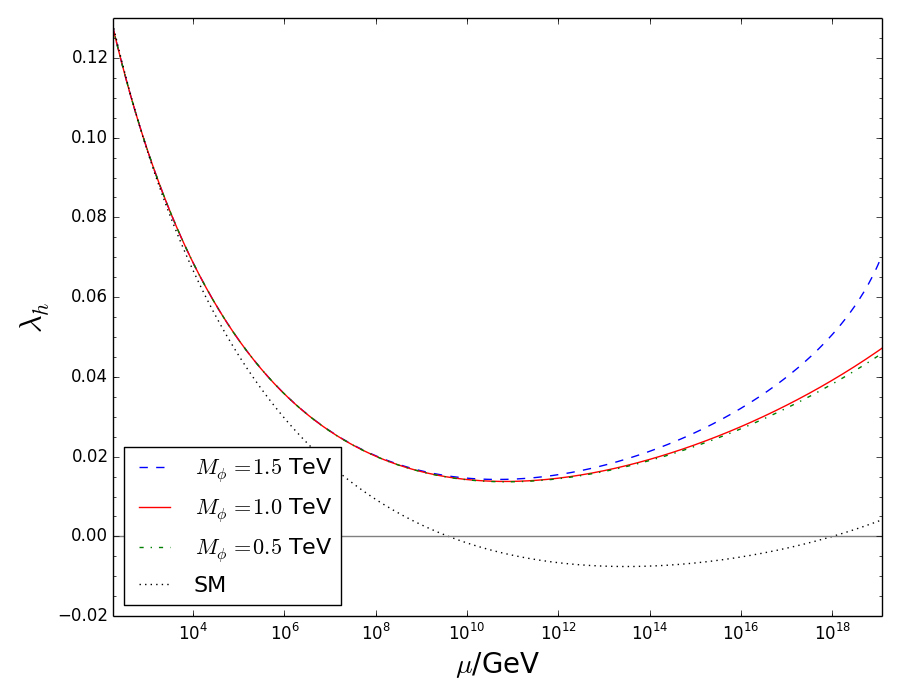}} \\
	\subfigure[$M_\phi = 1$ TeV, $\tan\theta = 0.1$, $\lambda_\phi(\Lambda)$ = 0.01]
	{\label{fig:lam_hphi} %
		\includegraphics[width=8.3cm]{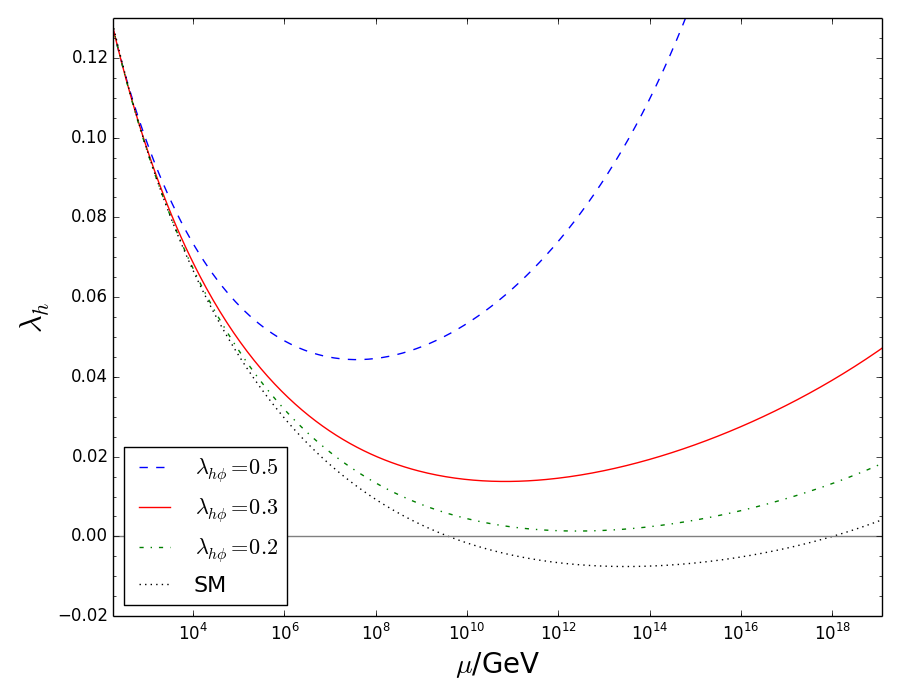}} \quad
	\subfigure[$M_\phi = 1$ TeV, $\tan\theta = 0.1$, $\lambda_{h\phi}(\Lambda)$ = 0.3]
	{\label{fig:lam_phi} %
		\includegraphics[width=8.3cm]{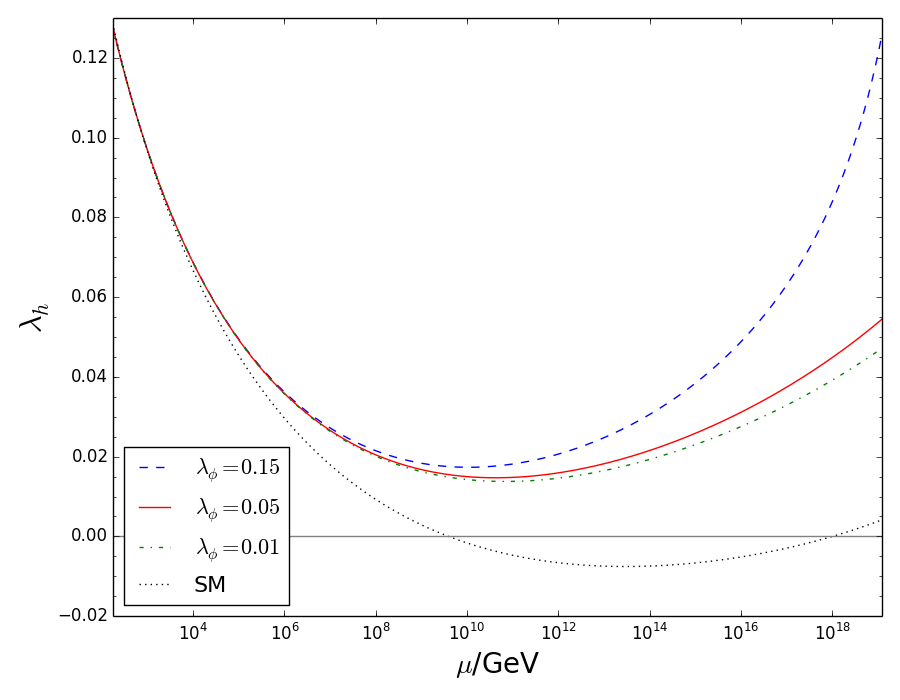}}	
	\caption{Running of $\lambda_h$ for different values of $M_\phi$, $\tan\theta$, $\lambda_{h\phi}(\Lambda)$,
		and $\lambda_{\phi}(\Lambda)$. }
	\label{fig:lambda_h}
\end{figure}

The metastability of the electroweak vacuum can be overcome by the extended Higgs sector 
with portal couplings to the Higgs.  
The new scalar couplings can change the $\beta$ function and the infrared boundary condition
of the Higgs quartic.
The $\beta$ function of a coupling $\lambda_i$ at a scale $\mu$ in the renormalization group equation is defined as
$\beta_{\lambda_i} = \partial \lambda_i/\partial \log\mu$.
For dimensionless couplings in the scalar potential (including the SM Yukawa coupling),
the one-loop $\beta$ functions are given by 
\beqr \label{eq:betafunction}
\beta_{\lambda_{t}}^{(1)} &=& \frac{\lambda_{t}}{16\pi^2}\left[ \frac{9}{2}\lambda_t^2 
-\left(8g_3^2 + \frac{9}{4}g_2^2 + \frac{17}{12}g_1^2\right) \right],
\nnb \\[1pt]
\beta_{\lambda_h}^{(1)} &=& \frac{1}{16\pi^2}\left[ 24\lambda_h^2 + 12\lambda_h\lambda_t^2 -6\lambda_t^4
-3\lambda_h(3g_2^2 + g_1^2) +\frac{3}{8}\big(2g_2^4+(g_2^2+g_1^2)^2\big) + \half\lambda_{hs}^2 
+ \frac{N}{2}\lambda_{h\phi}^2 \right],
\nnb \\[1pt]
\beta_{\lambda_{hs}}^{(1)} &=& \frac{1}{16\pi^2}\left[ \lambda_{hs}\left(12\lambda_h 
-\frac{3}{2}(3g_2^2 + g_1^2) + 6\lambda_t^2 + 4\lambda_{hs} + 6\lambda_s\right) 
+ N\lambda_{h\phi}\lambda_{s\phi}  \right],
\nnb \\[1pt]
\beta_{\lambda_s}^{(1)} &=& \frac{1}{16\pi^2}\left[2\lambda_{hs}^2 + 18 \lambda_s^2 
+ \frac{N}{2}\lambda_{s\phi}^2 \right], 
\nnb \\[1pt]
\beta_{\lambda_{h\phi}}^{(1)} &=& \frac{1}{16\pi^2}\left[ \lambda_{h\phi}\left(12\lambda_h 
-\frac{3}{2}(3g_2^2 + g_1^2) + 6\lambda_t^2 + 4\lambda_{h\phi} + 2(N+2)\lambda_\phi\right) 
+ \lambda_{hs}\lambda_{s\phi} \right],
\nnb \\[1pt]
\beta_{\lambda_{s\phi}}^{(1)} &=& \frac{1}{16\pi^2}\bigg[\lambda_{s\phi}\bigg(4\lambda_{s\phi}
+ 6\lambda_s + 2(N+2)\lambda_\phi \bigg) + 4\lambda_{hs}\lambda_{h\phi} \bigg], 
\nnb \\[1pt]
\beta_{\lambda_{\phi}}^{(1)} &=& \frac{1}{16\pi^2}\left[2\lambda_{h\phi}^2 + 2(N+8) \lambda_\phi^2 
+ \half \lambda_{s\phi}^2  \right], 
\eeqr 	
where $g_1$ and $g_2$ are the SM U$(1)_Y$ and SU$(2)_L$ couplings, respectively,
and $\lambda_t$ is the top Yukawa coupling.
While we consider the NP effects on the effective potential and the $\beta$ functions at one-loop order,
in order to show how much the NP contribution is needed for stabilizing the scalar potential,
we include the following two-loop $\beta$ functions for the top-Yukawa and Higgs quartic self-couplings
as done in Ref.~\cite{Baek:2012uj} because those contributions are sizable:
\beqr
\beta_{\lambda_{t}}^{(2)} &\simeq& \frac{\lambda_{t}}{(16\pi^2)^2}
\bigg[-12\lambda_t^4-12\lambda_t^2\lambda_h+6\lambda_h^2
	+\lambda_t^2\left(36g_3^2+\frac{225}{16}g_2^2+\frac{131}{16}g_1^2\right)
\nnb \\[1pt]
	&& +g_3^2\left(9g_2^2+\frac{19}{9}g_1^2\right) -108g_3^4 -\frac{3}{4}g_2^2g_1^2
	-\frac{23}{4}g_2^4 + \frac{1187}{216}g_1^4 \bigg],
\nnb \\[1pt]
\beta_{\lambda_h}^{(2)} &\simeq& \frac{1}{(16\pi^2)^2}
\bigg[ \lambda_h\lambda_t^2\left(-144\lambda_h-3\lambda_t^2 
	+ 80g_3^2 + \frac{85}{6}g_1^2 + \frac{45}{2}g_2^2 \right)
\nnb \\[1pt]
&& + \lambda_h\left(-312\lambda_h^2 + \lambda_h\left(36g_1^2 + 108g_2^2\right)
	\frac{629}{24}g_1^4 - \frac{73}{8}g_2^4 +\frac{39}{4}g_1^2+g_2^2 \right)
\nnb \\[1pt]
&& + \lambda_t^2\left( 30\lambda_t^4 -\lambda_t^2\left(\frac{8}{3}g_1^2 + 32g_3^2\right)  
-\frac{19}{4}g_1^4 - \frac{9}{4}g_2^4 + \frac{21}{2}g_1^2g_2^2 \right)	
\nnb \\[1pt]
&& + \frac{1}{48}\left(915g_2^6 - 379g_1^6 - 289g_1^2g_2^4 - 559g_1^4g_2^2 \right)	
\bigg].
\eeqr
The $\beta$ functions for the SM gauge couplings are not altered by the NP couplings up to the next-leading order 
and can be found in Ref.~\cite{Schrempp:1996fb}.
For numerical simulation, we assume the central values for the top and gauge boson masses
and use the following SM values:
$g_1(M_t) = 0.464,\, g_2(M_t) = 0.648,\, g_3(M_t) = 1.167,\, \lambda_t (M_t) = 0.951,\, \lambda_h (M_t) = 0.129$.	
As one can see from Eq.~(\ref{eq:betafunction}),
the portal coupling $\lambda_{hs}$ and the SM-DM interaction coupling $\lambda_{h\phi}$
give positive contributions to the $\beta$ function of the Higgs quartic.
Especially for large $N$, $\lambda_{h\phi}$ contributions are much enhanced.
Also, the contribution of $\lambda_{\phi}$ to the $\beta$ function of $\lambda_{h\phi}$ is sizable,
so the DM self-coupling is also important in stabilizing the scalar potential.

\begin{figure}[!hbt]
	\centering%
	\includegraphics[width=8.3cm]{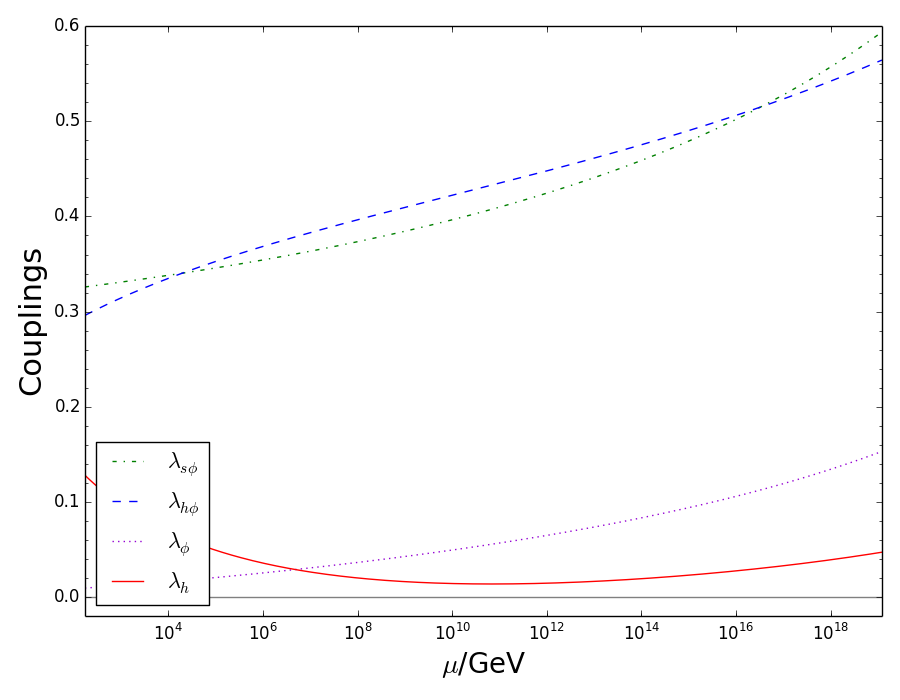} \quad
	\includegraphics[width=8.3cm]{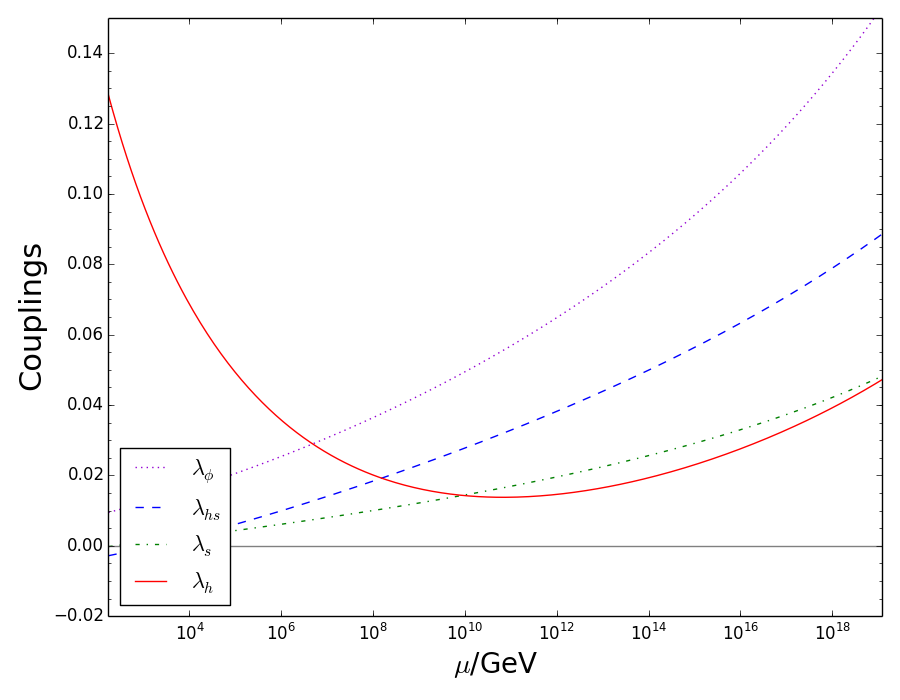} 
	\caption{Running of the scalar couplings for 
		$M_\phi = 1$ TeV, $\tan\theta = 0.1$, $\lambda_{h\phi}(\Lambda)$ = 0.3, 
		$\lambda_\phi(\Lambda)$ = 0.01.}
	\label{fig:lambdas}
\end{figure}

In Fig.~\ref{fig:lambda_h}, we plot the running of the Higgs quartic coupling $\lambda_{h}$ 
for different values of $M_\phi$, $\tan\theta$, $\lambda_{h\phi}(\Lambda)$, and $\lambda_{\phi}(\Lambda)$ in the case of $N =2$
with which the scalar vacua are stable and all scalar couplings are perturbative (less than 4$\pi$) up to the Planck scale. 
One can clearly see that the nonzero coupling $\lambda_{h\phi}$ plays very important role in stabilizing the Higgs potential.
Also in Fig.~\ref{fig:lambdas}, the running behavior of the other dimensionless scalar couplings is shown 
for the benchmark points of the new parameters: 
$M_\phi = 1$ TeV, $\tan\theta = 0.1$, $\lambda_{h\phi}(\Lambda)$ = 0.3, $\lambda_\phi(\Lambda)$ = 0.01. 
We scanned all new parameter space satisfying the vacuum stability and the perturbativity of the couplings   
and will apply this result to the DM phenomenology study in the next section.

\section{Phenomenology}

We first consider the relic density constraints on this model. 
At present, the most accurate determination of the DM mass density $\Omega_{\rm DM}$ 
comes from global fits of cosmological parameters to a variety of observations
such as Planck primary cosmic microwave background (CMB) data plus the Planck measurement of CMB lensing
 \cite{ParticleDataGroup:2020ssz}:
\beq
\Omega_{\rm CDM}h^2 = 0.1200 \pm 0.0012.
\label{eq:relic_obserb}
\eeq
This relic density observation will exclude some regions in the model parameter space.
The relic density analysis in this section includes all possible channels of $\phi_i\phi_i$ pair annihilation into the SM particles.  	
Using the numerical package micrOMEGAs \cite{Belanger:2018mqt} 
that utilizes CalcHEP for computing the relevant annihilation cross sections \cite{Belyaev:2012qa}, 
we compute the DM relic density and the spin-independent DM-nucleon scattering cross sections.
Especially, micrOMEGAs is known to be effective for the relativistic treatment of the thermally averaged cross section
and for a precise computation of the relic density in the region where annihilation 
through a Higgs exchange occurs near resonance \cite{Belanger:2004yn}. 

\begin{figure}[!hbt]
	\centering%
	\subfigure[ $\lambda_{\phi} = 0.01$ ]{\label{lamhpr1} %
		\includegraphics[width=8.6cm]{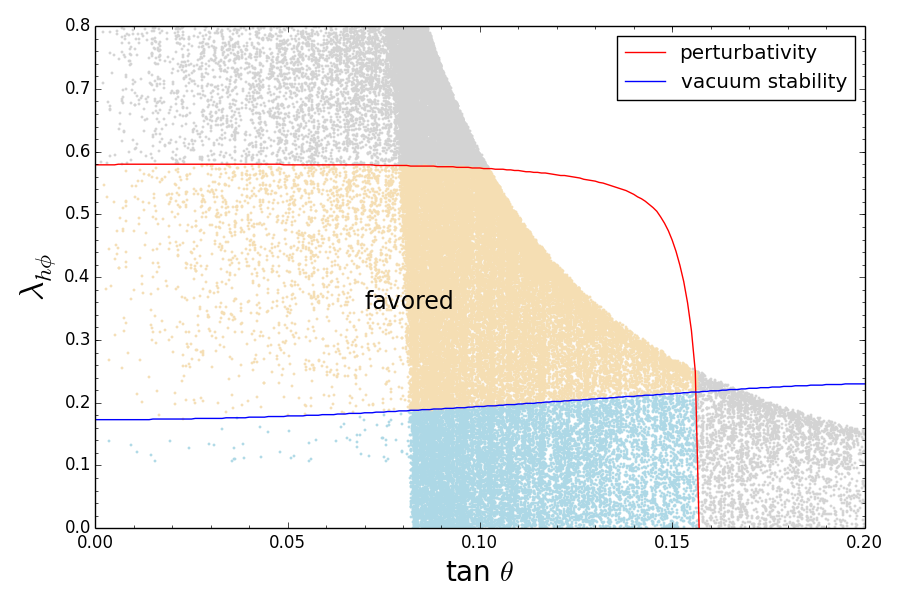}} \
	\subfigure[ $\lambda_{\phi} = 0.1$ ]{\label{lamhpr2} %
		\includegraphics[width=8.6cm]{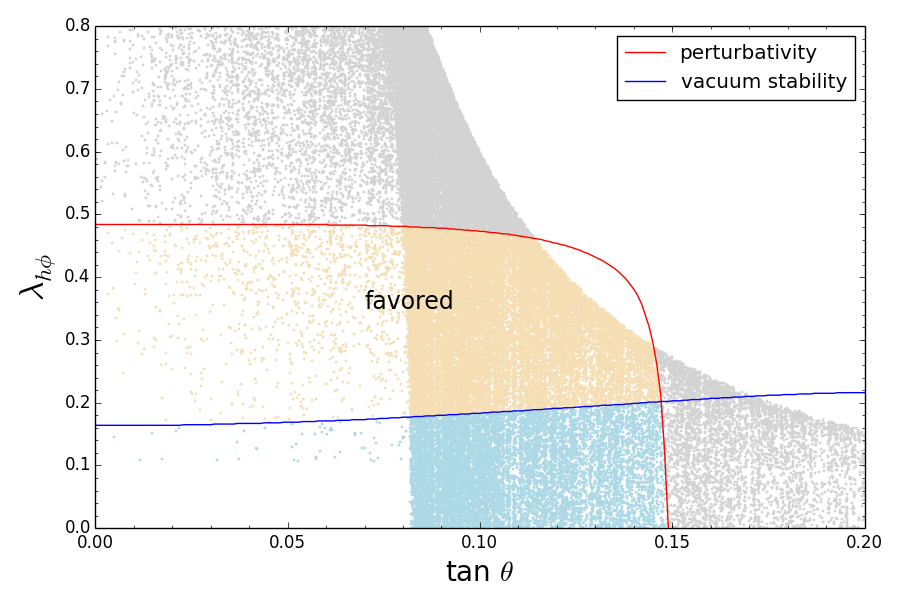}}   	
	\caption{Allowed regions for the parameter sets of ($\lambda_{h\phi}, \tan\theta$) 
		by relic density observations at $3\sigma$ level for $M_2 \leq 800$ GeV. 
	    The gold region satisfies both of the vacuum stability and perturbativity conditions obtained in Sec.~III,
	    the light blue region satisfies only the perturbativity condition, and the remaining data are grayed out.      
    } 
	\label{fig:vevstability}
\end{figure}

\begin{figure}[!hbt]
	\centering%
	\subfigure[ $\lambda_{\phi} = 0.01$]{\label{omegahr1} %
		\includegraphics[width=8.6cm]{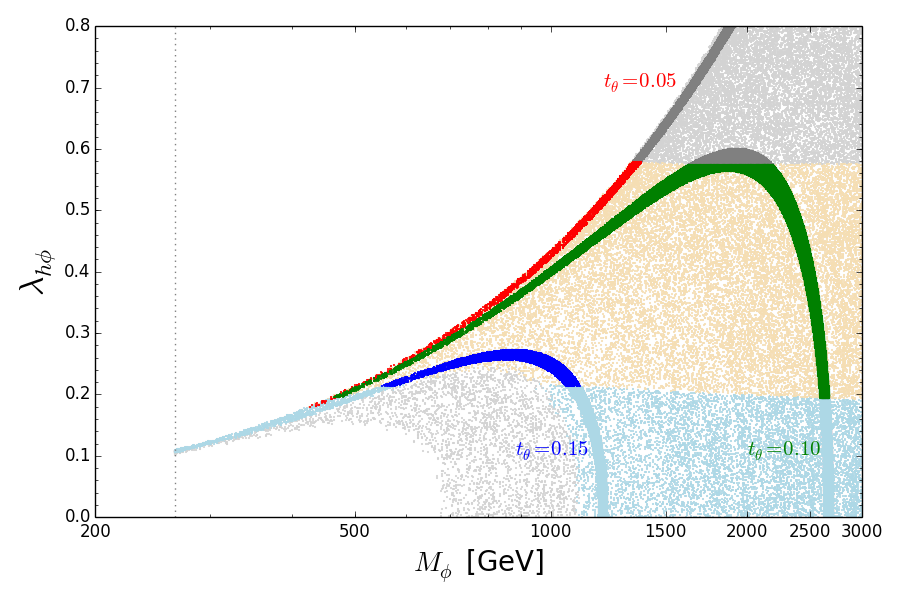}} \
	\subfigure[ $\lambda_{\phi} = 0.1$]{\label{omegahr2b} %
		\includegraphics[width=8.6cm]{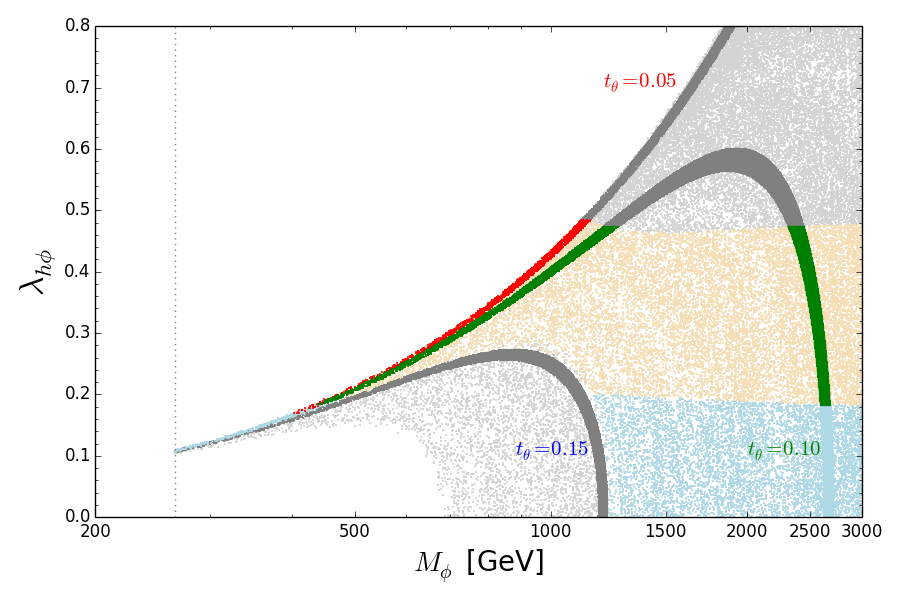}}     
	\caption{Allowed regions for the parameter sets of ($M_\phi, \lambda_{h\phi}$) 
		by relic density observations at $3\sigma$ level for $0 \leq \tan\theta \leq 0.2$ and $M_2 \leq 800$ GeV. 
	    The gold region satisfies both of the vacuum stability and perturbativity conditions obtained in Sec.~III,
        the light blue region satisfies only the  perturbativity condition, and the remaining data are grayed out.
        For reference, three benchmark points for the scalar mixing angle $\tan\theta = 0.05, 0.10, 0.15$ are represented 
        as red, green, and blue, respectively, and
        the vertical dotted line indicates the minimum value of $M_\phi$ obtained in Eq.~(\ref{eq:radit_mass})
        for $N=2$.
	} 
	\label{fig:relicdensity}
\end{figure}

For illustration of allowed new model parameter spaces, we choose the benchmark points 
for the DM self-coupling $\lambda_{\phi} = 0.01, 0.1$.
We consider the $N=2$ case only for simplicity, 
which corresponds to the case containing two exact copies of the DM. 
Large values of $N$ are disfavored because this ruins perturbativity of the scalar couplings at high scale.
Using the conditions provided in Sec.~III, 
we perform the phenomenological analysis of the model by varying the following three NP
parameters: $\tth$, $M_{\phi}$, $\lambda_{h\phi}$. 
The new scalar mass $M_2$ is determined by $M_\phi$ and $\tth$ from Eq.~(\ref{eq:radit_mass}),
and the dependency of other NP parameters are shown in Eq.~(\ref{eq:parameters}).
To see the relic density constraints on the scalar DM interaction,
we first plot the allowed region of the Higgs-DM coupling $\lambda_{h\phi}$ and the scalar mixing angle $\tan\theta$ 
constrained by the current relic density observations at 3$\sigma$ level for $M_2 \leq 800$ GeV in Fig.~\ref{fig:vevstability}.
If one adopts the conditions that lead to a strong first-order phase transition as needed to
produce the observed baryon asymmetry of the Universe, the preferred $h_2$ mass should be less than 1 TeV \cite{Profumo:2007wc}.
The chosen upper bound on $M_2$ corresponds to the several TeV of $M_\phi$ depending on the value of $\tan\theta$.
In the figure, the red (blue) line corresponds to the upper (lower) bound of perturbativity (vacuum stability) condition.
The gold and light blue colored data satisfy the perturbativity condition obtained in Sec.~III,
and the remaining data are grayed out.	
The gold and light blue data represent the stable and metastable vacua, respectively. 
As emphasized earlier, one can clearly see that $\lambda_{h\phi}$ should be sizable 
in order to satisfy all the theoretical and phenomenological constraints. 
The larger value of $\lambda_{\phi}$ gets a stronger bound from the perturbativity constraint
	because it contributes to the $\beta$ function of the Higgs quartic positively,
	so the smaller value of $\lambda_{\phi}$ derives the larger allowed parameter space. 
Figure~\ref{fig:relicdensity} shows the allowed regions of the DM mass $M_\phi$ and $\lambda_{h\phi}$ 
by relic density observations at $3\sigma$ level for $0 \leq \tan\theta \leq 0.2$ similarly to Fig.~\ref{fig:vevstability}. 
For reference, three benchmark points for the scalar mixing angle $\tan\theta = 0.05, 0.10, 0.15$ are represented 
as red, green, and blue, respectively, in order to see $\tan\theta$ dependence clearly.
The behavior of the favored $\lambda_{h\phi}$ for a fixed $\theta$ was explained in much detail 
including relevant formulas in our previous study \cite{Jung:2019dog},
so we will not repeat it here.  
Although $\lambda_{\phi}$ was assumed to be zero in the previous study,
the nonzero $\lambda_{\phi}$ case in this work shows similar behavior but with stronger constraints. 

\begin{figure}[!hbt]
	\centering%
	\subfigure[ $\lambda_{\phi} = 0.01$]{\label{sigmaSIr1} %
		\includegraphics[width=8.6cm]{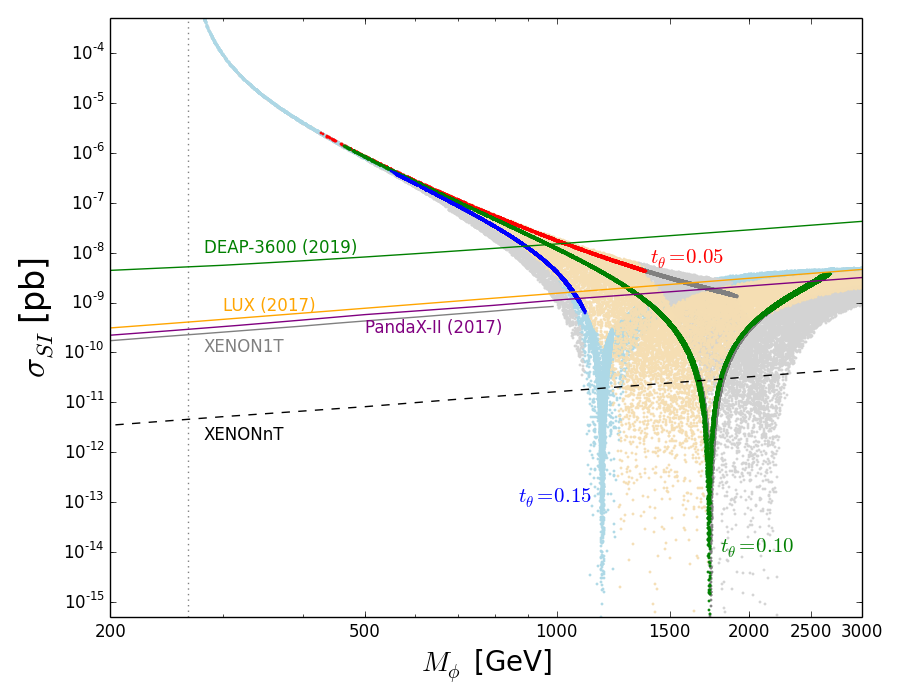}} \
	\subfigure[ $\lambda_{\phi} = 0.1$]{\label{sigmaSIr2} %
		\includegraphics[width=8.6cm]{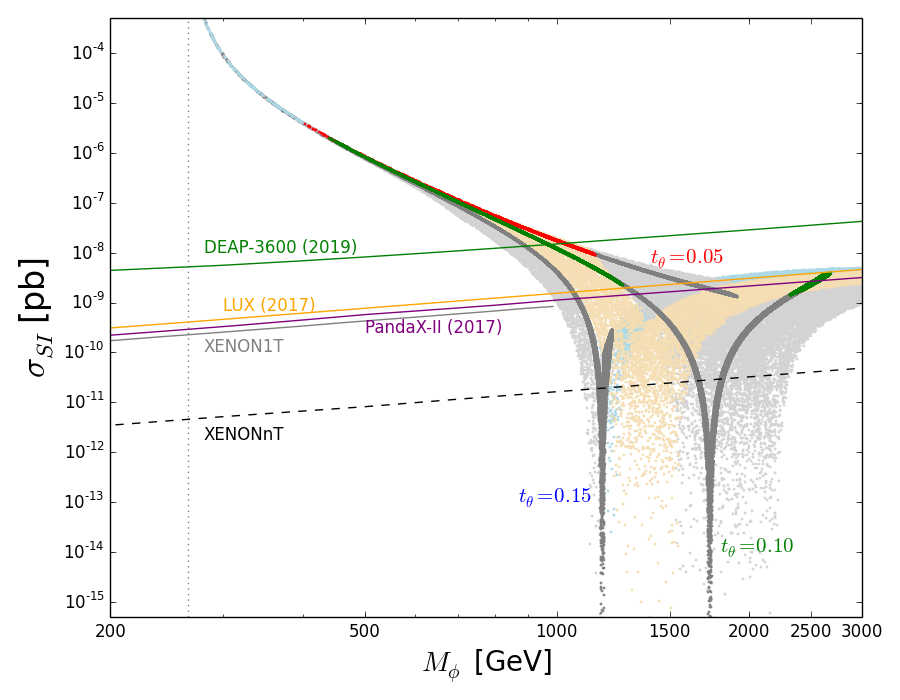}}  	 
	\caption{Spin-independent DM-nucleon scattering results allowed by relic density observations
		for $0 \leq \lambda_{h\phi} \leq 0.8$ and $0 \leq \tth \leq 0.2$.  
		Also shown are observed limits from DEAP-3600 (2019), LUX (2017), PandaX-II (2017), XENON1T    
		and expected limits from XENONnT. 
       The gold region satisfies both of the vacuum stability and perturbativity conditions obtained in Sec.~III,
       the light blue region satisfies only the perturbativity condition, and the remaining data are grayed out.
       For reference, three benchmark points for the scalar mixing angle $\tan\theta = 0.05, 0.10, 0.15$ are represented
       as red, green, and blue, respectively, and
       the vertical dotted line indicates the minimum value of $M_\phi$ obtained in Eq.~(\ref{eq:radit_mass})
       for $N=2$.
	}
	\label{fig:sigmaSI}
\end{figure}

Next, we consider the implications of the direct detection experiments on this model. 
Nonobservation of DM-nucleon scattering events is interpreted as an upper bound on DM-nucleon cross section. 
In Fig.~\ref{fig:sigmaSI}, 
we plot the spin-independent DM-nucleon scattering cross section by varying the DM mass $M_\phi$ 
with parameter sets allowed by the relic density observation within $3\sigma$ range 
and compare the results with the observed upper limits
obtained at 90\% level from DEAP-3600 (2019) \cite{DEAP:2019yzn}, LUX (2017) \cite{LUX:2016ggv}, 
PandaX-II (2017) \cite{PandaX-II:2017hlx}, XENON1T \cite{XENON:2018voc}   
and expected limits from XENONnT \cite{XENON:2020kmp}.
The DM-nucleon scattering occurs only through the two $t$-channel diagrams exchanging $h_1$ and $h_2$. 
If $h_2$ is very light, then
the enhancement of the $h_2$ exchange diagram dominates 
over the contribution of the SM-like Higgs boson $h_1$.
As we increase the DM mass, $M_2$ and $\lambda_{s\phi}$ increase as well,
and $h_1$ contributions become more significant. 
The cross sections hit those minima when $M_2 \simeq M_1$ 
and increase again mainly due to $\lambda_{s\phi}$. 
One can see from the figure that the allowed mixing angle $\tan\theta$ is highly constrained in the heavy $M_\phi$ region
and our lower bound on the DM mass is $M_\phi \sim$ 1070 GeV. 

There are other observational constraints on DM annihilation cross section 
such as Fermi-LAT \cite{Ackermann:2015zua} and  H.E.S.S. \cite{Abdallah:2016ygi} measurements.
They give stringent limits on the annihilation cross sections of the DM, especially on $b\bar{b}$ and $\tau\bar{\tau}$
for a lighter DM mass less than a few hundred GeV.
However, our model prefers DM mass heavier than about 1 TeV 
and predicts far smaller cross sections than the experimental constraints.
Also, there is an observational constraint on the extra Higgs portal scalar particle. 
Big bang nucleosynthesis gives a constraint on the lifetime of the scalar particle $h_2$ 
less than 1 sec \cite{Jedamzik:2009uy,Kaplinghat:2013yxa}, and it is consistent in our model.  

Since the hidden sector is connected to the SM by the Higgs portal in this model,
there are also constraints from the collider experiments. 
Our choice of the mixing angle $t_\theta \leq 0.2$ is quite safe against the LEP2 constraints 
since the $h_2$ mass exceeds the corresponding LEP2 lower mass bound for the DM mass over $1$ TeV. 
Additional constraints from nonobservation of Higgs-like particles 
in the high-mass Higgs searches through $WW$ and $ZZ$ modes
at the LHC \cite{GonzalezLopez:2016xly,CMS:2016ilx,Angelidakis:2017ebh} 
do not give severe restrictions to our analysis. 
A similar discussion on the Higgs portal scalar with a fermionic DM can be found in Ref.~\cite{Kim:2019ogz}.
For future collider experiments, 
the perturbativity bound on $\tth < 0.156$ obtained in Fig.~\ref{fig:vevstability} in this model
does not lie within the expected precision of VLHC and HL-LHC on $\tth$ 
from the deviation of experimental value of the Higgs self-coupling $c_{111}$ \cite{Dawson:2013bba}.
But the deviation of the experimental value of Higgs boson couplings lies 
within the expected precision of ILC experiment in $ZZ$ modes for $\tth \gtrsim 0.12$ at $\sqrt{s} = 250$ GeV
and for $\tth \gtrsim 0.11$ at $\sqrt{s} = 500$ GeV \cite{Barklow:2017suo}.
Therefore, $\tth \gtrsim 0.11$ case in this model will be able to be tested in future collider experiments sooner or later.

\section{Summary and Conclusion}

In this work, we investigated vacuum structure and vacuum stability 
in an extension of the SM which is renormalizable and classically scale invariant.
We introduced the SM gauge singlet DM sector that consists of a real scalar field S 
as a pseudo-Nambu–Goldstone boson of scale symmetry breaking
and a scalar multiplet of global ${\cal O}(N)$ symmetry as a viable DM candidate.
The communication between the SM and the singlet DM sectors is accomplished by the Higgs portal interaction. 
The scalar masses are generated quantum mechanically through the Coleman-Weinberg mechanism for the EWSB. 
Also, the DM scalar $\phi$ couples directly to the SM Higgs with the coupling $\lambda_{h\phi}$
which plays an important role in resolving the vacuum stability issue as well as in DM phenomenology.  

Adopting the framework of GW, we chose a flat direction at tree level lifted by radiative corrections. 
Through the mixing of the scalar mediator with the SM-like Higgs boson, 
two light scalar particles $h_1$ and $h_2$ interact with both visible and hidden sectors. 
After EWSB, the SM Higgs $h_1$ and the DM scalars $\phi$ have the tree level masses, 
while the new scalar singlet $h_2$ acquires its mass through radiative corrections
of the SM particles and $\phi$ as obtained in Eq.~(\ref{eq:radit_mass})
using the relationship between Higgs portal couplings simplified at the GW scale.
The metastability of the electroweak vacuum can be overcome by the new dimensionless couplings,   
which can change the $\beta$ function and the infrared boundary condition of the Higgs quartic.

With four independent new parameters $\tth$, $M_\phi$, $\lambda_{h\phi}$, and $\lambda_{\phi}$, 
we presented the allowed region of NP parameters by the relic density observation 
satisfying the vacuum stability and the perturbativity constraints
in Figs.~\ref{fig:vevstability} and \ref{fig:relicdensity} for the $N=2$ case.
We also show and the spin-independent DM-nucleon scattering cross section of the scalar DM 
by varying the DM mass $M_\phi$ with the same allowed parameter sets 
and compare the results with the observed upper limits from various experiments in Fig.~\ref{fig:sigmaSI}.
In the figures, one can clearly see that
the allowed parameter space constrained by both the relic density observation and the vacuum stability condition is located 
in the mass region of $M_\phi$ heavier than about 1 TeV.
We performed the numerical analysis for $\tth \leq 0.2$, 
and found that the $\tth \leq 0.05$ case is disfavored in this model due to the recent PandaX-II bound.
We also found that our model is not constrained by the current indirect detection bounds for the given parameter sets,
but the $\tth \gtrsim 0.11$ case will be able to be tested in future collider experiments such as ILC.

\acknowledgments
This work was supported by Basic Science Research Program through the National Research Foundation of Korea (NRF)
funded by the Ministry of Education under the Grants No. 
NRF-2020R1I1A1A01072816 (S.-h.~Nam) and No. NRF-2021R1F1A1061717 (Y.~G.~Kim) 
and also funded by the Ministry of Science and ICT under the Grants No. 
NRF-2020R1A2C3009918 (S.-h.~Nam and J.~Lee) and No. NRF-2021R1A2C2011003 (K.~Y.~Lee).

\def\npb#1#2#3 {Nucl. Phys. B {\bf#1}, #2 (#3)}
\def\plb#1#2#3 {Phys. Lett. B {\bf#1}, #2 (#3)}
\def\prd#1#2#3 {Phys. Rev. D {\bf#1}, #2 (#3)}
\def\jhep#1#2#3 {J. High Energy Phys. {\bf#1}, #2 (#3)}
\def\jpg#1#2#3 {J. Phys. G {\bf#1}, #2 (#3)}
\def\epj#1#2#3 {Eur. Phys. J. C {\bf#1}, #2 (#3)}
\def\arnps#1#2#3 {Ann. Rev. Nucl. Part. Sci. {\bf#1}, #2 (#3)}
\def\ibid#1#2#3 {{\it ibid.} {\bf#1}, #2 (#3)}
\def\none#1#2#3 {{\bf#1}, #2 (#3)}
\def\mpla#1#2#3 {Mod. Phys. Lett. A {\bf#1}, #2 (#3)}
\def\pr#1#2#3 {Phys. Rep. {\bf#1}, #2 (#3)}
\def\prl#1#2#3 {Phys. Rev. Lett. {\bf#1}, #2 (#3)}
\def\ptp#1#2#3 {Prog. Theor. Phys. {\bf#1}, #2 (#3)}
\def\rmp#1#2#3 {Rev. Mod. Phys. {\bf#1}, #2 (#3)}
\def\zpc#1#2#3 {Z. Phys. C {\bf#1}, #2 (#3)}
\def\cpc#1#2#3 {Chin. Phys. C {\bf#1}, #2 (#3)}

\end{document}